\documentclass[preprint2]{aastex}

\slugcomment{submitted to ApJ}

\shorttitle{Star Formation and X-ray Sources}
\shortauthors{Kaaret and Alonso-Herrero}

\begin{document}

\title{X-Ray Sources in the Star Forming Galaxies NGC 4194 and NGC 7541}

\author{Philip Kaaret\altaffilmark{1,2} and Almudena
Alonso-Herrero\altaffilmark{3}}  

\altaffiltext{1}{Department of Physics and Astronomy, University of
Iowa, Van Allen Hall, Iowa City, IA 52242, USA}

\altaffiltext{2}{Universit\'e Paris 7 Denis Diderot and Service
d'Astrophysique, UMR AIM, CEA Saclay, F-91191 Gif sur Yvette, France}

\altaffiltext{3}{Departamento de Astrofisica Molecular e Infrarroja,
Instituto de Estructura de la Materia, CSIC, E-28006 Madrid, Spain}


\begin{abstract}

We examine the X-ray point source population and 2-10~keV luminosity for
two galaxies with high star formation rates (SFRs), NGC 4194 and NGC
7541.  The X-ray point source luminosity function (XLF) for these two
galaxies is consistent with the XLF found by Grimm et al.\ (2003) for a
sample of star-forming galaxies.  Combining our results with a sample of
galaxies with SFRs above 1~$M_{\odot}$/yr, we find that the number of
X-ray point sources above a luminosity of $2 \times 10^{38} \rm \, erg
\, s^{-1}$ is $N = (1.8 \pm 0.4) {\rm SFR}/(M_{\odot} \rm \, yr^{-1})$. 
This number is lower than previously inferred by Grimm et al.\ based on
a sample of galaxies with lower SFRs.  We find that the ratio of X-ray
luminosity in the 2-10~keV band to SFR is $L_X/(10^{40} \rm \, erg \,
s^{-1}) = (0.37 \pm 0.08) {\rm \, SFR}/(M_{\odot} \rm yr^{-1})$.  This
value may serve as a calibration in attempts to use X-ray luminosity to
measure the SFR of galaxies at cosmological distances.  The ratio of
mass accreted onto compact objects versus mass used to form stars is
near $10^{-6}$.  This ratio may be useful in constraining population
synthesis models of X-ray binary formation in actively star forming
systems.

\end{abstract}

\keywords{black hole physics -- galaxies: individual: NGC 4194 --
galaxies: individual: M100 -- galaxies: individual: NGC 4945 --
galaxies: individual: NGC 7541 -- galaxies: stellar content -- X-rays:
galaxies}

\section{Introduction}

The bulk of bright X-ray sources in most galaxies are compact objects
formed at the endpoint of the stellar evolution of massive stars
(typically $> 8 M_{\odot}$).  Because compact objects can be visible as
X-ray binaries for lifetimes far exceeding those of their progenitor
massive stars, the X-ray binary populations may be used as `fossils' to
study the star formation history of their host galaxies
\citep{Griffiths90,David92}.  Recently, \citet{Ranalli03} demonstrated a
correlation between the 2-10 keV X-ray luminosity and star formation
rates (SFR) for 23, mainly nearby, galaxies.  

\citet{Grimm03}, hereafter GGS03, extended these results to consider the
X-ray point source luminosity function (XLF) in addition to the total
X-ray luminosity.  They found a correlation between the number of X-ray
sources in a galaxy and the star formation rate.  This correlation does
not suffer the strong Malmquist bias of the luminosity correlation.
However, due to the limitations of their sample, GGS03 measured the
correlation between the number of X-ray sources in a galaxy and the star
formation rate only for galaxies with star formation rates of $7
M_{\odot}/\rm yr$ or less.  It is of interest to extend these
measurements to galaxies with larger star formation rates.

We have obtained {\it Chandra} observations of two galaxies, NGC 4194
and NGC 7541, selected from the sample of \citet{Devereux89}.  Both have
high star formation rates, near $12 M_{\odot}/\rm yr$, but are close
enough, nearer than 40~Mpc, to distinguish individual X-ray sources.  
See Table~\ref{galaxies} for some basic properties of the galaxies.  NGC
4914, also known as the Medusa, is a minor merger at a distance of
39.5~Mpc \citep{Weistrop04}.  NGC 7541 is an SBc galaxy and has a
companion galaxy NGC 7537.   It was the host of SN1998dh which allows an
accurate distance estimate of $37.1 \pm 1.5$~Mpc \citep{Jha07}.  We also
extracted results from the literature and from the data archives for
additional galaxies with star formation rates above $1 M_{\odot}$/yr.
Our observations and analysis are described in section \S 2.  In \S 3,
we discuss the X-ray point source luminosity functions.  In \S 4, we
discuss the relation of X-ray source populations to star formation rate.

\begin{table}[tb]
\begin{center}
\begin{tabular}{lll}
\hline
Name                     & NGC 4194      & NGC 7541     \\ \hline
RA (J2000)               & 12 14 09.6    & 23 14 43.86  \\
DEC (J2000)              & +54 31 35.8   & +04 32 01.8  \\
Distance (Mpc)           & 39.5          & 37.1         \\
SFR ($M_{\odot}$/yr)     & 12.3          & 12.0         \\
Major diameter           & 1.82$\arcmin$ & 3.47$\arcmin$ \\
Minor diameter           & 1.10$\arcmin$ & 1.15$\arcmin$ \\
Position angle           & 164$\arcdeg$  & 102$\arcdeg$  \\ \hline
X-ray sources            & 5-10          & 15           \\
$L_{X} (10^{40} \rm \, erg \, s^{-1})$        
                         & 3.0           & 1.8          \\ \hline
\end{tabular} \end{center}

\caption{contains for each galaxy: the galaxy name, RA, DEC, distance,
star formation rate (SFR), the major and minor diameters and position
angle of the $d_{25}$ ellipse from the RC3 catalog, the number of  X-ray
sources above a luminosity of $2 \times 10^{38} \rm \, erg \, s^{-1}$,
and the total X-ray luminosity $(L_{X})$ in the 2-10 keV band.  The
distances are from \citet{Weistrop04} for NGC 4914 and from
\citet{Jha07} for NGC 7541.}  \label{galaxies} \end{table}

\section{Observations and Analysis}

\subsection{New observations}

Observations of NGC 4194 and NGC 7541 were made using the Advanced CCD
Imaging Spectrometer (ACIS; Bautz et al.\ 1998) on board the Chandra
X-Ray Observatory.  The {\it Chandra\/} observation of NGC 4194 (ObsID
7071) began on 9 September 2006 at 23:30:18 UT and had a useful exposure
of 35.5~ks.  The observation of NGC 7541 (ObsID 7070) began on 31
October 2006 at 21:35:08 UT and had a useful exposure of 39.0~ks.  The
Chandra data were subjected to standard data processing and event
screening.  The total rate on the S3 chip was below 1.5~c/s for NGC 4194
and below 1.6~c/s for NGC 7541 indicating that there were no strong
background flares.

For each observation, we constructed an image using all valid events  in
the 0.3--8~keV band on the S3 chip and used the {\it wavdetect} tool
which is part of the {\it CIAO} version 3.4 data analysis package to
search for X-ray sources.  For NGC 4194, which has extended diffuse
X-ray emission near its core, we also searched for sources in the
1.5--8~keV band.  Lists of sources with detection significance of
$3.0\sigma$ or higher within the $d_{25}$ ellipse of each galaxy are
given in Tables~\ref{ngc4194src} and \ref{ngc7541src}.  In order to
calculate the source fluxes, we computed an exposure map for an assumed
source spectrum of a powerlaw with photon index of 1.7 absorbed by  the
total Galactic {\sc H i} column density along the line of sight
\citep{Dickey90} which we take as $5.2 \times 10^{20} \rm \, cm^{-2}$
for NGC 4194 and $1.5 \times 10^{20} \rm \, cm^{-2}$ for NGC 7541.   We
also calculated source luminosities in the 2-10~keV band using the same
spectral model and assuming that the sources are at the same distance as
the galaxy. In order to give some indication of the spectral shape of
each source, we calculate the ratio of the Chandra counts in the
1--8~keV band to counts in the 0.3--8~keV band.  The source fluxes,
luminosities, and colors are listed in Tables~\ref{ngc4194src} and
\ref{ngc7541src}.

\begin{deluxetable}{rllrrcccl}
\tabletypesize{\scriptsize}
\tablecaption{X-ray sources within the field of NGC 4194
  \label{ngc4194src}}
\tablewidth{0pt}
\tablehead{ \# & \colhead{RA} & \colhead{DEC} & 
  \colhead{S/N} & \colhead{Counts} &  
  \colhead{Flux} & \colhead{Luminosity} & \colhead{Hardness}  &
  \colhead{Counterpart} \\
	& & & & & ($10^{-15} \, \rm erg \, cm^{-2} \, s^{-1}$) & ($10^{38} \, \rm erg \, s^{-1}$) }
\startdata
  1 & 12 14 09.69 & +54 32 15.6 & 164.8 &  455 & 110 $\pm$ 5   & 129  & 0.54 \\
  2 & 12 14 09.62 & +54 31 35.9 & 139.8 &  870 & 210 $\pm$ 7   & 246  & 0.68 & Nucleus \\
  3 & 12 14 06.18 & +54 31 43.0 & 126.2 &  390 &  95 $\pm$ 5   & 111  & 0.60 & SDSS \\
  4 & 12 14 09.99 & +54 31 27.1 &  13.3 &   86 &  22 $\pm$ 2   & 25.5 & 0.48 \\
  5 & 12 14 09.90 & +54 31 42.2 &   5.6 &   21 & 5.1 $\pm$ 1.1 &  5.9 & 0.95 \\
  6 & 12 14 09.23 & +54 31 44.9 &   4.7 &   21 & 5.1 $\pm$ 1.1 &  5.9 & 0.52 \\
\enddata
\tablecomments{Listed for each source: number; RA and DEC -- the
position of the source in J2000 coordinates; S/N -- significance of the
source detection as calculated by {\it wavdetect}; Counts -  counts in
the 0.3--8~keV band; Flux -- source flux in units of $10^{-15} \, \rm
erg \, cm^{-2} \, s^{-1}$ in the 0.3--8~keV band calculated assuming a
power law spectrum with photon index of 1.7 and corrected for the
Galactic absorption column density along the line of sight; Luminosity
-- source luminosity in units of $10^{38} \, \rm erg \, s^{-1}$ in the
2--10~keV band using the same spectral model; Hardness - the hardness
ratio defined as the ratio of counts in the 1--8~keV band to counts in
the 0.3--8~keV band.}   \end{deluxetable}

\begin{deluxetable}{rllrrcccl}
\tabletypesize{\scriptsize}
\tablecaption{X-ray sources within the field of NGC 7541
  \label{ngc7541src}}
\tablewidth{0pt}
\tablehead{ \# & \colhead{RA} & \colhead{DEC} & 
  \colhead{S/N} & \colhead{Counts} &  
  \colhead{Flux} & \colhead{Luminosity}  & \colhead{Hardness}  &
  \colhead{Counterpart} \\
	& & & & & ($10^{-15} \, \rm erg \, cm^{-2} \, s^{-1}$) & ($10^{38} \, \rm erg \, s^{-1}$) }
\startdata
  1 & 23 14 42.21 & +04 31 40.1 &  71.0 &  229 &   49 $\pm$ 3   & 57.0 & 0.73 \\
  2 & 23 14 42.39 & +04 32 31.0 &  44.7 &  121 &   26 $\pm$ 2   & 30.3 & 0.73 \\
  3 & 23 14 38.85 & +04 32 04.1 &  23.0 &   63 & 13.6 $\pm$ 1.7 & 15.8 & 0.73 \\
  4 & 23 14 43.91 & +04 32 02.6 &  22.3 &  141 &   30 $\pm$ 3   & 35.1 & 0.60 \\
  5 & 23 14 39.17 & +04 32 04.3 &  20.4 &   50 & 10.8 $\pm$ 1.5 & 12.5 & 0.80 \\
  6 & 23 14 39.81 & +04 32 18.2 &  19.3 &   46 &  9.9 $\pm$ 1.5 & 11.6 & 0.96 \\
  7 & 23 14 46.22 & +04 31 56.4 &  14.9 &   48 & 10.3 $\pm$ 1.5 & 12.0 & 0.83 \\
  8 & 23 14 39.92 & +04 32 02.0 &  13.9 &   39 &  8.4 $\pm$ 1.3 &  9.7 & 0.62 \\
  9 & 23 14 46.01 & +04 32 03.2 &  11.4 &   36 &  7.7 $\pm$ 1.3 &  9.0 & 0.89 \\
 10 & 23 14 46.97 & +04 31 54.2 &   8.3 &   21 &  4.5 $\pm$ 1.0 &  5.2 & 0.76 \\
 11 & 23 14 42.63 & +04 32 15.7 &   6.5 &   18 &  3.9 $\pm$ 0.9 &  4.5 & 0.78 \\
 12 & 23 14 40.21 & +04 32 03.9 &   5.8 &   14 &  3.0 $\pm$ 0.8 &  3.5 & 1.00 \\
 13 & 23 14 42.43 & +04 32 03.4 &   4.7 &   13 &  2.8 $\pm$ 0.8 &  3.2 & 0.77 & 2MASS \\
 14 & 23 14 47.33 & +04 32 06.4 &   4.6 &   10 &  2.2 $\pm$ 0.7 &  2.5 & 0.80 \\
 15 & 23 14 41.64 & +04 31 51.7 &   3.4 &    8 &  1.7 $\pm$ 0.6 &  2.0 & 0.88 \\
\enddata

\tablecomments{The quantities are as defined for
Table~\ref{ngc4194src}.}   \end{deluxetable}

We searched for counterparts to the X-ray sources in the USNO B1 catalog
\citep{usnob1} to attempt to improve the {\it Chandra} astrometry.  We
found three matches within $1\arcsec$ in the S3 image for NGC 4194. 
After a shift of $0.3\arcsec$, the X-ray and optical positions for all
three objects agree within $0.22\arcsec$.  For NGC 7541, we found two
coincidences, but both were at the edge of the S3 chip where the {\it
Chandra} point spread function is relatively large and the source
location accuracy is relatively poor.  Thus, we chose not to correct the
astrometry for NGC 7541.  We also searched for counterparts in the 2MASS
catalog \citep{2mass}.  The only counterpart in the NGC 4194 field was
for the nucleus of the galaxy.  The core of NGC 4194 has strong,
extended near-IR emission which may obscure point sources near the
center of the galaxy.  There were several counterparts within $1\arcsec$
within NGC 7541, but crowding of 2MASS sources in the disk of the galaxy
prevents confident assignment of unique counterparts. 

We searched for X-ray variability by  comparing the photon arrival times
for each source to the distribution expected for a constant source with
the same average flux using a Kolmogorov-Smirnoff (KS) test.  No
background subtraction was performed. We find significant variability
only for CXOU J231438.85+043204.1 in NGC 7541.  This source shows a
marked decrease in flux across the observation.

\begin{deluxetable}{llcccc}
\tabletypesize{\scriptsize}
\tablecaption{X-ray spectra for bright point sources
  \label{xspec}}
\tablewidth{0pt}
\tablehead{
  \colhead{RA} & \colhead{DEC} & \colhead{$\chi^2$/DoF} & 
    \colhead{$\Gamma$} & \colhead{$N_H$} & \colhead{Flux} \\
  & & & & 
	  \colhead{(10$^{21}$ cm$^{-2}$)} & 
	  \colhead{(10$^{-14}$ erg cm$^{-2}$ s$^{-1}$)} }
\startdata
\multicolumn{6}{l}{NGC 4194} \\
12 14 09.69 & 54 32 15.6 & 18.2/36 & $2.6^{+0.5}_{-0.4}$ & $1.1^{+0.9}_{-0.6}$ & 6.1 \\
12 14 06.18 & 54 31 43.0 & 10.5/33 & $2.1^{+0.5}_{-0.4}$ & $0.6^{+0.8}_{-0.1}$ & 6.3 \\
\hline
\multicolumn{6}{l}{NGC 7541} \\
23 14 42.21 & 04 31 40.1 & 13.8/19 & $2.2^{+0.6}_{-0.5}$ & $2.6^{+1.6}_{-1.2}$ & 3.3 \\
23 14 42.39 & 04 32 31.0 &  2.3/12 & $2.0^{+0.9}_{-0.7}$ & $2.0^{+2.2}_{-1.8}$ & 2.1 \\
23 14 43.91 & 04 32 02.6 &  8.1/11 & $2.3^{+1.3}_{-0.9}$ & $1.8^{+2.6}_{-1.6}$ & 1.5 \\
\enddata

\tablecomments{Listed for each source: the position in RA and DEC
(J2000), the goodness of the fit, the photon index, the absorption
column density, and the absorbed flux in the 0.3--8~keV band.}
\end{deluxetable}

We performed X-ray spectral fitting for the sources with more than 100
counts.  We fitted the spectra using the {\it Sherpa} fitting package
and response matrices calculated using the {\tt mkrmf} tool in {\it
CIAO}.  We used the $\chi^2$-Gehrels statistic to evaluate the quality
of the fits due to the low numbers of counts.  The fit results are shown
in Table~\ref{xspec}.  All of the spectra, except for the nuclear region
of NGC 4194, were adequately fitted with an absorbed power-law.  NGC
4194 shows extended X-ray emission near the nucleus, so we extracted two
spectra: one in a circular region with a radius of 2 pixels and the
other with a radius of 13 pixels covering the region of extended
emission.  We fit these spectra with a model consisting of a power-law
plus thermal emission from diffuse gas, specifically the APEC model. 
The large region spectrum is best fitted, $\chi^2/{\rm DoF} = 54.3/99$,
when the thermal component has a  temperature $kT = 0.77 \pm 0.06$~keV
and an unabsorbed flux of $(9.7 \pm 1.8) \times 10^{-14} \rm \, erg \,
cm^{-2} \, s^{-1}$ in the 0.3--8~keV band, the power-law component has a
flux of $(2.3 \pm 0.6) \times 10^{-13} \rm \, erg \, cm^{-2} \, s^{-1}$
and a photon index $\Gamma = 2.0 \pm 0.3$, and the absorption column
density is $N_H = (1.3 \pm 0.5) \times 10^{21} \rm \, cm^{-2}$.  We 
fitted the small region spectrum with the APEC temperature fixed to $kT
= 0.77$.  In the best fit, $\chi^2/{\rm DoF} = 22.0/33$, the power-law
component describing the nuclear point source had a photon index $\Gamma
= 1.7 \pm 0.4$ and an unabsorbed flux of $(9 \pm 3) \times 10^{-14} \rm
\, erg \, cm^{-2} \, s^{-1}$ in the 0.3--8~keV band where we have
corrected for the encircled energy within the 2 pixel radius.

\begin{figure*}[tb]
\centerline{\includegraphics[width=6.0in]{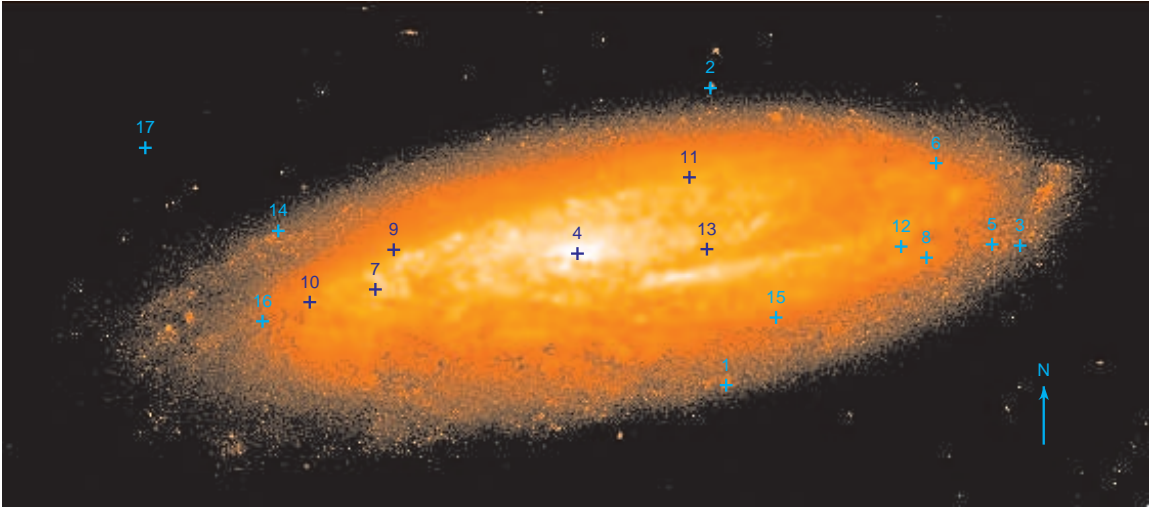}} \caption{HST/ACS
F814W (I-band) image of NGC 7541.  The crosses indicate the positions of
the X-ray sources listed in Table~\ref{ngc7541src}.  The arrow points
North and has a length of $10\arcsec$.} \label{i7541} \end{figure*}

Observations of both galaxies were made using the Advanced Camera for
Surveys (ACS) on the Hubble Space Telescope (HST) under GO program
10769.  Images were obtained in the broad band filter F814W (I-band)
using the Wide-Field Camera (WFC).  NGC 7541 was observed on 8 August
2006.  We used the images delivered by the standard ACS pipeline
processing (OPUS 2006\_5 and CALACS code version 4.6.1) which removes
cosmic-rays, corrects for optical distortion, and dither combines the
images.  We aligned the image to stars in the USNO B1.0 catalog
\citep{usnob1} using the Graphical Astronomy and Image Analysis Tool
({\it GAIA}; Starlink version 2.8-0).  Fig.~\ref{i7541} shows the
positions of the X-ray sources detected with Chandra overlaid on the
HST/ASC I-band image of NGC 7541.

\begin{figure}[tb] \centerline{\includegraphics[width=3.0in]{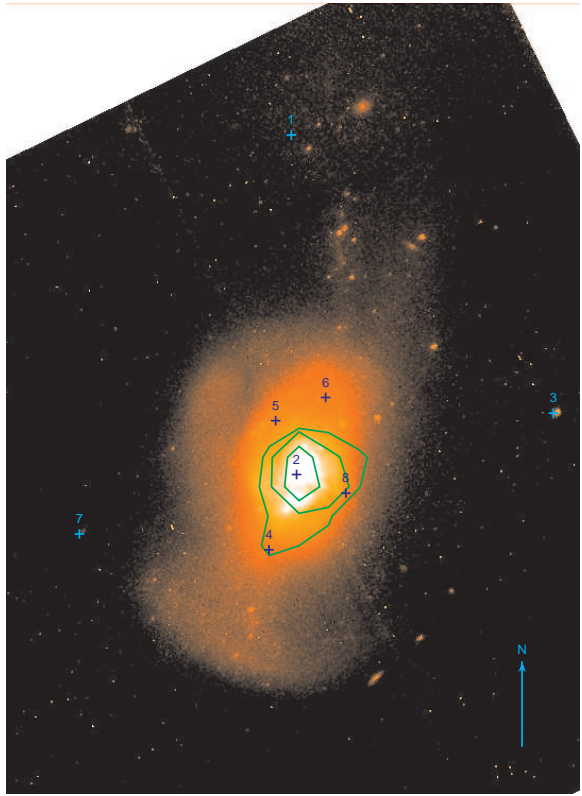}}
\caption{NICMOS F110W infrared image of NGC 4194.  The green contours
represent the diffuse X-ray emission and correspond to 0.5, 1, and 2
counts per ACIS pixel ($0.492\arcsec$ square).  The diffuse X-ray
emission is located at the region with a dense stellar population as
indicated by the IR intensity.  The crosses indicate the positions of
the X-ray sources listed in Table~\ref{ngc4194src}.  The arrow points
North and has a length of $10\arcsec$.} \label{nic4194} \end{figure}

Observations of both galaxies were also made with the Near Infrared
Camera and Multi-Object Spectrometer (NICMOS) on HST using the NIC3
camera that has a pixel size of $0.2\arcsec$ and a field of view (FOV)
of $\sim 51.2\arcsec \times 51.2\arcsec$.  We used two broad-band
filters: F110W and F160W.  The images were reduced using the {\it
HST}/NICMOS pipeline routines, which involve subtraction of the first
readout, dark current subtraction on a readout-by-readout basis,
correction for linearity and cosmic ray rejection, and flat fielding. 
Since the size of the galaxies exceeds the FOV of the NIC3 camera, for
each galaxy we took four pointings to produce a final mosaic covering an
approximate FOV of $1.6\arcmin \times 1.6\arcmin$.  We aligned the
mosaic images to stars in the 2mass catalog \citep{2mass} using {\it
GAIA}.  Fig.~\ref{nic4194} shows the positions of the X-ray sources and
contours of the X-ray diffuse emission overlaid on the F110W mosaic of
NGC 4194.

One of the sources in the NGC 4194 field with an USNO B1 counterpart
falls within the $d_{25}$ ellipse of the galaxy.  The R magnitude is
19.43. The X-ray to optical flux ratio of the source, $\log(f_x/f_R) =
\log f_X + 5.50 +R/2.5 = 0.9$, is in the range found for AGN
\citep{Hornschemeier01}.   This source also has a Sloan Digital Sky
Survey counterpart, SDSS J121406.21+543143.1, which is identified as a
faint quasar based on the optical colors.  The source lies within the
$d_{25}$ ellipse of NGC 4194, but not in the main body of the galaxy or
the tidal tail.  On the HST F814W (I-band) image, we find no stars close
to the source other than the counterpart.  Thus, we conclude that the
source is most likely a background AGN.

\subsection{Additional galaxies}

We analyzed archival Chandra data for the galaxies M100 and NGC 4945
following the same procedures described above.  The observation of NGC
4945 was made on 27 Jan 2000 (ObsID 864) beginning at 19:00:10 UT with
an exposure of 24.5~ks.  We reprocessed the data using recent
calibrations and procedures current in Ciao 3.4.  The galaxy is larger
than the ACIS S3 chip, but the roll angle of the observation placed the
entire $D_{25}$ ellipse of the galaxy on ACIS chips S2, S3, and S4. 
Thus, we extracted sources and calculated exposure maps for all three
chips.  The observation of M100 was made on 18 Feb 2006 (ObsID 6727)
beginning at 19:00:10 UT with an exposure of 24.5~ks.  The target was SN
1979C, which is not at the center of M100, so we included ACIS chips S2
and S3 in order to cover the whole galaxy.  Lists of sources with
detection significance of $3.0\sigma$ or higher within the $d_{25}$
ellipse of each galaxy are given in Tables~\ref{m100src} and
\ref{ngc4945src}.  We estimated the number of background AGN at fluxes
above that equivalent to $2 \times 10^{38} \rm \, erg \, s^{-1}$
within each galaxy using the source counts from \citet{Giacconi01} in
the 2--10~keV band.

\begin{table}[tb]
\begin{center}
\begin{tabular}{lccc}
\hline
Galaxy          & Distance &  SFR             & Sources \\
                & (Mpc)    & ($M_{\odot}$/yr) &         \\ \hline
NGC 4194        & 39.5     & 12.3             & 10      \\
NGC 7541        & 37.1     & 12.0             & 15      \\
NGC 1068        & 14.4     &  9.0             & 18      \\
M51             &  8.4     &  4.9             & 13      \\
Antennae        & 13.8     &  3.6             &  9      \\
M100            & 16.8     &  3.3             & 11      \\
NGC 4945        &  3.7     &  2.7             &  4      \\
M82             &  3.6     &  1.4             &  7      \\
NGC 4579        & 16.8     &  1.3             &  3      \\
M74             &  8.8     &  1.2             &  2      \\ \hline
\end{tabular} \end{center}

\caption{contains for each galaxy: the galaxy name, distance, star
formation rate, and net number of X-ray sources above a luminosity of $2
\times 10^{38} \rm \, erg \, s^{-1}$ in the 2-10 keV band after removal
or subtraction of background AGN.} \label{nsfr_table} \end{table}

For the galaxies M51, M74, M82, NGC 4579, NGC 1068, and the Antennae, we
extracted source counts from the literature.  In each case, we 
estimated the number of background AGN included in the count using the
source counts from \citet{Giacconi01} unless this was already done in
the reference or background and foreground sources were already excluded
by a search for counterparts at other wavelengths.  Below we will
integrate the XLF to find the total X-ray luminosity due to point
sources.  For this to be valid, the energy band used to find the point
source luminosities must match the energy band used to calculate the
total X-ray luminosity, specifically 2--10~keV.  Thus, we corrected the
source luminosities from the literature to this energy band.  We used
the spectral model assumed in each paper.  However, using a power-law
model as for NGC 4194 and NGC 7541 produced no significant change.

We note that GGS03 extracted X-ray point source luminosities from the
literature and made no attempt to correct for the differing energy bands
and assumptions concerning spectra shape made in the various papers. We
find that, for the galaxies in their sample, making the energy band
correction produces a large change only for the Antennae.  In addition
to the energy band correction, \citet{Ivo04} recently found a distance
to the Antennae based on the red giant branch tip of $13.8 \pm 1.7$~Mpc
that is significantly lower than previous distances including that used
by GGS03.  Also, one of the Antennae sources was identified as a
background AGN \citep{Clark05}.  Thus, the star formation rate for the
Antennae becomes $3.6 \, M_{\odot} \rm \, yr^{-1}$ and the number of
sources above $2 \times 10^{38} \rm \, erg \, s^{-1}$ becomes 9.

For M51, we find 11 net sources above $2 \times 10^{38} \rm \, erg \,
s^{-1}$ in the 2-10~keV band using the 2--8 keV band results of
\citet{Terashima04}.  For M74, we estimate two sources from the 2--8 keV
band results of \citet{Kilgard02,Kilgard05}.  For M82, we estimate 7 net
sources from the 2--10 keV band results of \citet{Griffiths00}.  For NGC
4579, we estimate 3 sources from the results of \citet{Eracleous02}. 
For NGC 1068, we estimate 18 net sources from the 0.5--8 keV band
results of \citet{Smith03}.  For these galaxies, we have adopted the
SFRs quoted by GGS03 rescaled to the distances listed in
Table~\ref{nsfr_table} except for NGC 1068 and NGC 4945 where we have
calculated the SFR from the FIR luminosities quoted in
\citet{Koribalski96} rescaled to the distances quoted in
Table~\ref{nsfr_table}.  Both are Seyfert 2 galaxies containing an
active galactic nucleus which produces infrared emission.  Thus, the SFR
calculated from the total FIR luminosity may overestimate the true SFR. 
However, the starburst in NGC 1068 is known to dominate the flux at
wavelengths longer than 30~$\mu$m \citep{Telesco84}, so AGN
contamination does not strongly affect the SFR estimate.

\begin{figure}[tb] \centerline{\includegraphics[width=3.25in]{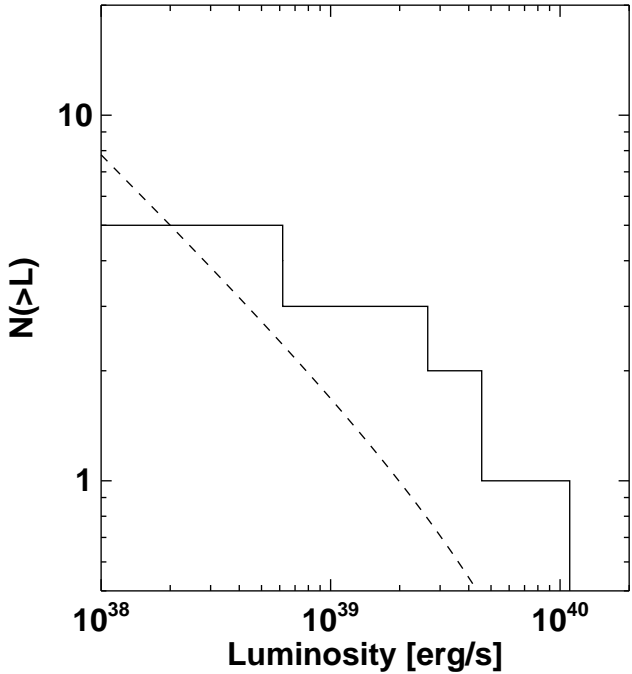}} 
\caption{\label{xlf4194} X-ray point source luminosity function for NGC
4194.  The dashed line indicates the functional form from GGS03
normalized to have the same number of sources with luminosity above $2
\times 10^{38} \rm \, erg \, s^{-1}$ as observed in the galaxy.}
\end{figure}

\begin{figure}[tb] \centerline{\includegraphics[width=3.25in]{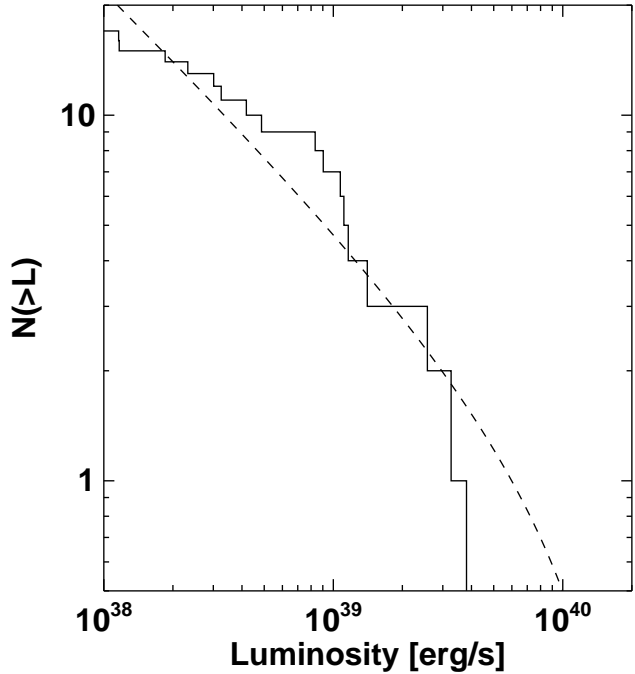}}
\caption{\label{xlf7541} X-ray point source luminosity functions for NGC
7541.  The dashed line indicates the functional form from GGS03
normalized to have the same number of sources with luminosity above $2
\times 10^{38} \rm \, erg \, s^{-1}$ as observed in the galaxy.}
\end{figure}

\section{X-ray point source luminosity functions}

Figs.~\ref{xlf4194} and \ref{xlf7541} show the X-ray point source
luminosity functions (XLFs) for NGC 7541 and NGC 4194.  The dashed
curves are the `universal' XLFs for star-forming galaxies suggested by
GGS03.  The dashed curves are normalized to have the same number of
sources with luminosity above $2 \times 10^{38} \rm \, erg \, s^{-1}$ as
observed in the comparison galaxy.  A Kolmogorov-Smirnov test indicates
that the distributions for both galaxies are consistent with this form.

At low luminosities, $2-5 \times 10^{38} \rm \, erg \, s^{-1}$, the XLF
for  NGC 4194 appears flatter than the GGS03 XLF.   The observation
duration is sufficient to detect sources with luminosities down to at
least $2 \times 10^{38} \rm \, erg \, s^{-1}$ with high efficiency;  a
luminosity of $2 \times 10^{38} \rm \, erg \, s^{-1}$ should produce 7
counts and a $3.2\sigma$ detection.  However, the diffuse X-ray emission
near the nucleus, where most of the star formation is concentrated,
hampers our ability to detect sources.  Indeed, there is a possible
source located at $\rm R.A. = 12^h  14^m 08\fs96$, $\rm Decl. =
+54\arcdeg 31\arcmin 33\farcs7$ (J2000) that lies in the region of
diffuse emission.  The source is marginally detected at a significance
of $2.4\sigma$ and only in the 1.5--8~keV band, but appears to have a
luminosity of $5 \times 10^{38} \rm \, erg \, s^{-1}$.

\begin{figure}[tb] \centerline{\includegraphics[width=3.25in]{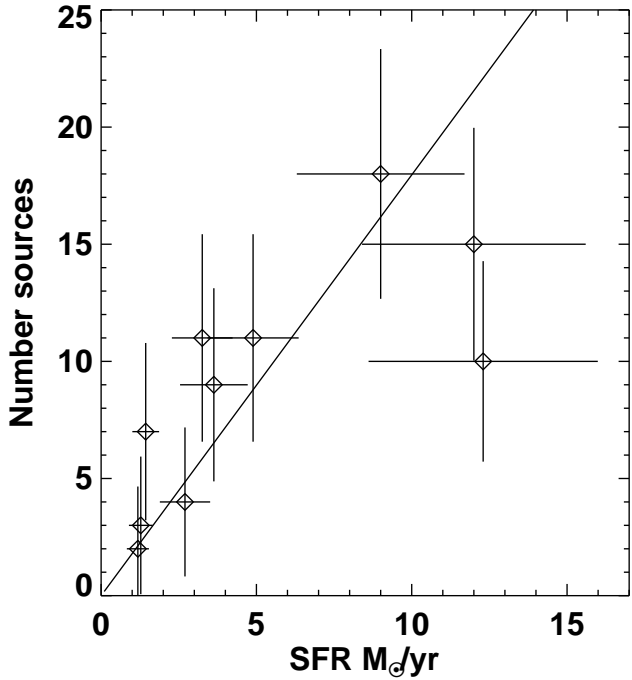}}
\caption{\label{nsfr} Number of point sources with luminosity above $2
\times 10^{38} \rm \, erg \, s^{-1}$ versus star formation rate for the
galaxies listed in Table~\ref{nsfr_table}.  The solid line is the linear
fit described in the text.}  \end{figure}

The excess 2-10 keV flux in the nuclear region not accounted for in
detected point sources is $5.5 \times 10^{-14} \rm \, erg \, cm^{-2} \,
s^{-1}$ equivalent to a luminosity of $1.0 \times 10^{40} \rm \, erg \,
s^{-1}$.  By integrating a differential XLF, we can find the number of
sources above a given luminosity.  By weighting the integrand by the
luminosity, we can find the total luminosity produced by the set of
sources above a given luminosity.  For the GGS03 XLF, specifically eq.~6
in their paper, the relation between the number of sources, $N$, with
luminosities above $2 \times 10^{38} \rm \, erg \, s^{-1}$ and the total
luminosity, $L$, of all point sources is $N = 4.9 L/10^{40} \rm \, erg
\, s^{-1}$.  If the excess emission in the central region arises from
point sources, then $N = 5.0^{+1.7}_{-1.2}$ sources would be expected. 
Using this estimate, the adjusted number of point sources in NGC 4194
would be 10.  This is the number used in Table~\ref{nsfr_table} and
Fig.~\ref{nsfr}.

\section{Relation of X-ray source population to star formation rate}

Our primary goal is to extend the study of the relation between star
formation rate and X-ray point source populations of galaxies to higher
star formation rates (SFRs).  NGC 7541 and NGC 4194 have SFRs
significantly higher than any galaxy for which the X-ray point source
population (as opposed to the total galactic X-ray luminosity) was
studied by GGS03.

For NGC 4194, using the FIR luminosity versus SFR conversions adopted in
GGS03 to calculate the SFR for stars more massive than $5M_{\odot}$ and
correcting for our assumed distance of 39.5~Mpc, we estimate a SFR of
$9.1 \, M_{\odot} \rm \, yr^{-1}$ based on the far infrared (FIR)
luminosity quoted by \cite{Buat02}, a rate of $12.6 \, M_{\odot} \rm \,
yr^{-1}$ based on the FIR luminosity in \citet{Young96}, and a rate of
$15.3 \, M_{\odot} \rm \, yr^{-1}$ based on the FIR luminosity in
\citet{Deutsch87}.  The 1.4~GHz radio flux of NGC 4194 is 91.0~mJy when
measured with a $1.5\arcsec$ beam \citep{Condon90}, which gives a SFR of
$30 \, M_{\odot} \rm \, yr^{-1}$ ($32.0 \, M_{\odot} \rm \, yr^{-1}$ for
a $1.5\arcsec$ beam) using the relations in GGS03.  The wide range of
the SFR estimates for NGC 4194 highlight the difficulties with SFR
estimation.  Because we compare the X-ray luminosity directly with the
FIR luminosity below, we adopt the mean, $12.3 \, M_{\odot} \rm \,
yr^{-1}$, of the three FIR estimates as our best estimate of the SFR.

For NGC 7541, we estimate a SFR of $12.0 \, M_{\odot} \rm \, yr^{-1}$
based on the FIR luminosity in \citet{Young96} calculated from the 60
and 100~$\mu$m fluxes, and a rate of $12.1 \, M_{\odot} \rm \, yr^{-1}$
based on the FIR luminosity calculated from the 12, 25, 60, and
100~$\mu$m fluxes from the IRAS survey using the formula in
\citet{Deutsch87}.  We adopt the mean, $12.0 \, M_{\odot} \rm \,
yr^{-1}$, of the two FIR estimates as our best estimate of the SFR.

Fig.~\ref{nsfr} shows the relation between star formation rate and the 
number of X-ray point sources with luminosities above $2 \times 10^{38}
\rm \, erg \, s^{-1}$ for the galaxies listed in
Table~\ref{nsfr_table}.  Fitting a linear relation to the data, we find
a slope of $N(L > 2 \times 10^{38} {\rm \, erg \, s^{-1}}) = (1.8 \pm
0.4) {\rm SFR}/(M_{\odot} \rm \, yr^{-1})$.  This fit was done
considering the errors in both the SFR and the number of sources.  The
error on the SFR was taken as 30\% of the SFR, for consistency with
GGS03.  The error on the number of sources was calculated using the
method of \citet{Gehrels86}.  The fit has $\chi^2/{\rm DoF} = 6.1/9$.

We note that our slope is lower than that quoted by GGS03.  We expect
that the number of sources found for each galaxy by GGS03 is higher than
the true number for the following reasons.  GGS03 did not correct the
source luminosities quoted in the literature in various energy bands
usually extending down to 0.5 or 0.1~keV to the 2-10~keV band used for
the SFR versus X-ray luminosity relation.  The GGS03 sample contained
mainly nearby and large angular diameter galaxies, and, thus, would
likely include a significant number of background AGN.  These were not
subtracted out.  The GGS03 sample may be biased to galaxies with
unusually large X-ray point source populations, since the early {\it
Chandra} observations of galaxies tended to be of those with high X-ray
fluxes.  Also, their sample contained many face-on galaxies.  The source
detection included counts in the soft X-ray band (usually down to
0.3~keV), and was, thus, most efficient for face-on galaxies where
absorption within the host is low.  This could be corrected by using
only counts in the 2--10~keV for source detection.  However, the
relatively low effective area of Chandra above 2~keV would increase the
detection threshold above $2 \times 10^{38} \rm \, erg \, s^{-1}$ in
many cases.  It is also possible that the difference reflects some
dependence of X-ray source formation on the properties of the host
galaxy.

The total X-ray luminosity has been suggested as a SFR indicator by
\citet{Ranalli03}.  We estimated the X-ray flux in the 2--10~keV band
for each galaxy by summing the counts in the 2--8 keV band for all the
point sources and then converting this sum to a total count rate and
then to an unabsorbed energy flux assuming a power-law spectrum with a
photon index $\Gamma = 1.7$ and given the absorption column density
along the line of sight to the galaxy.  For NGC 4194, we included all
the counts in the larger nuclear region, but excluded the likely
background AGN.  We find a luminosity in the 2--10 keV band of $3.0
\times 10^{40} \rm \, erg \, s^{-1}$ for NGC 4194 and  $1.8 \times
10^{40} \rm \, erg \, s^{-1}$ for NGC 7541.  We note that both galaxies
have sufficiently high SFR that they are in the linear regime of the
$L_X$-SFR relation.  The ratio of 2-10 keV luminosity, $L_{40}$, in
units of $10^{40} \rm \, erg \, s^{-1}$, to SFR, in units of $M_{\odot}
\rm yr^{-1}$, is $L_{40}/\rm SFR = 0.24$ for NGC 4194 and 0.15 for
NGC7541.  From the relation quoted above between SFR and the number of
X-ray sources, and assuming the X-ray luminosity function of GGS03, we
calculate a value of $0.37 \pm 0.08$.  This value is somewhat lower
than, but not inconsistent with, the ratio of 0.5 found by
\citet{Ranalli03}.  Comparing directly with the FIR luminosity, we find
$L_X = 1.6 \times 10^{-4} L_{\rm FIR}$.

It is possible to estimate the fraction of mass involved in star
formation which is accreted onto compact objects.  We define $f =
\dot{M}/{\rm SFR}$, where $\dot{M} = L_X/(\eta c^2)$ and $\eta$ is the
efficiency for conversion of matter into radiation.  If $\eta \sim 0.1$,
then $f = (1.8 \times 10^{-6}) L_{40}/SFR$, where $L_{40}$ is the
luminosity in units of $10^{40} \rm \, erg \, s^{-1}$ and $SFR$ is the
star formation rate in $M_{\odot}/\rm yr$.  Using the X-ray luminosity
of SFR ratio found above, we find $f \sim 7 \times 10^{-7}$.  Allowing a
bolometric correction for the X-ray luminosity would increase this value
by a factor of $\sim 3$.  This value should constrain population
synthesis models of the formation and evolution of X-ray binaries.  We
note that this value includes only the `prompt' accretion, i.e.\ that
occurring while the star formation is still active.  Additional
accretion may occur, notably in the formation of low-mass X-ray
binaries, well after active star formation has subsided.

\section*{Acknowledgments}

We thank the anonymous referee for comments which helped improve the
paper.  PK acknowledges partial support from Chandra grant CXC GO4-7075X
and STScI grant HST-GO-10769.  This research has made use of the VizieR
catalogue access tool, CDS, Strasbourg, France.  This research has made
use of the USNOFS Image and Catalogue Archive operated by the United
States Naval Observatory, Flagstaff Station.  This publication makes use
of data products from the Two Micron All Sky Survey, which is a joint
project of the University of Massachusetts and the Infrared Processing
and Analysis Center/California Institute of Technology, funded by the
National Aeronautics and Space Administration and the National Science
Foundation.


\begin{deluxetable}{rllrrcccl}
\tabletypesize{\scriptsize}
\tablecaption{X-ray sources within the field of M100
  \label{m100src}}
\tablewidth{0pt}
\tablehead{ \# & \colhead{RA} & \colhead{DEC} & 
  \colhead{S/N} & \colhead{Counts} &  
  \colhead{Flux} & \colhead{Luminosity} & \colhead{Hardness}  &
  \colhead{Counterpart} \\
	& & & & & ($10^{-15} \, \rm erg \, cm^{-2} \, s^{-1}$) & ($10^{38} \, \rm erg \, s^{-1}$) }
\startdata
  1 & 12 22 58.69 & +15 47 51.8 &  65.0 &  169 &   37 $\pm$ 3   &  7.6 & 0.62 & SN 1979C \\
  2 & 12 22 54.78 & +15 49 16.2 &  51.2 &  470 &   11 $\pm$ 5   & 21.6 & 0.66 & C3 \\
  3 & 12 22 54.15 & +15 49 12.3 &  42.6 &  203 &   46 $\pm$ 3   &  9.3 & 0.45 & C2 \\
  4 & 12 22 49.12 & +15 48 31.2 &  32.6 &   87 &   20 $\pm$ 2   &  4.1 & 0.79 \\
  5 & 12 22 54.86 & +15 49 18.1 &  28.2 &  205 &   46 $\pm$ 3   &  9.4 & 0.56 & C1 \\
  6 & 12 22 50.40 & +15 48 17.5 &  23.6 &   58 & 13.1 $\pm$ 1.7 &  2.7 & 0.78 \\
  7 & 12 22 46.20 & +15 48 49.6 &  20.0 &   55 & 13.2 $\pm$ 1.8 &  2.7 & 0.75 & C6? \\
  8 & 12 22 55.68 & +15 49 24.9 &  16.3 &   88 &   23 $\pm$ 2   &  4.7 & 0.65 \\
  9 & 12 22 58.24 & +15 48 59.3 &  16.2 &   39 &  8.8 $\pm$ 1.4 &  1.8 & 0.69 \\
 10 & 12 22 50.93 & +15 50 42.7 &  14.9 &   46 &   16 $\pm$ 2   &  3.2 & 0.78 \\
 11 & 12 22 58.38 & +15 49 18.9 &  14.7 &   38 &  8.6 $\pm$ 1.4 &  1.8 & 0.87 \\
 12 & 12 22 51.59 & +15 49 37.7 &  14.7 &   35 &   28 $\pm$ 5   &  5.7 & 0.74 \\
 13 & 12 22 54.25 & +15 49 44.2 &  14.0 &   30 & 18.9 $\pm$ 3.5 &  3.8 & 0.67 & C4 \\
 14 & 12 22 54.96 & +15 49 20.1 &  13.9 &  104 &   23 $\pm$ 2   &  4.8 & 0.40 & Nucleus \\
 15 & 12 22 57.16 & +15 48 57.4 &  12.6 &   30 &  6.8 $\pm$ 1.2 &  1.4 & 0.57 \\
 16 & 12 22 43.76 & +15 51 01.9 &  11.8 &   47 &   17 $\pm$ 3   &  3.5 & 0.91 \\
 17 & 12 22 43.22 & +15 51 04.7 &  11.6 &   46 &   17 $\pm$ 3   &  3.5 & 0.70 & SDSS\\
 18 & 12 22 44.54 & +15 49 28.0 &  11.5 &   27 &   15 $\pm$ 3   &  3.0 & 0.78 \\
 19 & 12 23 00.87 & +15 46 38.4 &  11.0 &   24 &  5.3 $\pm$ 1.1 &  1.1 & 0.33 & 2mass \\
 20 & 12 22 50.83 & +15 47 01.4 &  10.3 &   24 &  5.4 $\pm$ 1.1 &  1.1 & 0.88 \\
 21 & 12 22 51.11 & +15 48 59.5 &   8.9 &   26 &  5.9 $\pm$ 1.2 &  1.2 & 0.69 \\
 22 & 12 22 54.52 & +15 49 04.4 &   8.5 &   31 &  7.0 $\pm$ 1.3 &  1.4 & 0.65 \\
 23 & 12 22 47.11 & +15 49 13.1 &   8.0 &   24 &  5.7 $\pm$ 1.2 &  1.2 & 0.88 \\
 24 & 12 22 55.99 & +15 48 41.3 &   7.6 &   18 &  4.0 $\pm$ 0.9 &  0.8 & 0.89 \\
 25 & 12 22 58.40 & +15 47 26.5 &   7.4 &   17 &  3.9 $\pm$ 1.0 &  0.8 & 0.71 \\
 26 & 12 22 52.06 & +15 47 14.6 &   6.3 &   16 &  3.6 $\pm$ 0.9 &  0.7 & 0.88 \\
 27 & 12 22 54.73 & +15 48 03.6 &   6.2 &   15 &  3.3 $\pm$ 0.9 &  0.7 & 1.00 \\
 28 & 12 22 55.97 & +15 49 09.9 &   6.1 &   22 &  5.0 $\pm$ 1.1 &  1.0 & 0.68 \\
 29 & 12 22 56.83 & +15 46 30.9 &   5.9 &   15 &  3.3 $\pm$ 0.9 &  0.7 & 0.80 \\
 30 & 12 23 01.97 & +15 51 33.1 &   5.5 &   15 &  5.0 $\pm$ 1.3 &  1.0 & 0.73 \\
 31 & 12 22 57.39 & +15 48 19.8 &   5.4 &   13 &  2.9 $\pm$ 0.8 &  0.6 & 0.69 & [FFH96] C64? \\
 32 & 12 22 52.23 & +15 48 59.9 &   5.1 &   13 &  2.9 $\pm$ 0.8 &  0.6 & 0.69 \\
 33 & 12 22 50.00 & +15 48 28.2 &   5.1 &   16 &  3.6 $\pm$ 0.9 &  0.7 & 0.44 \\
 34 & 12 22 49.75 & +15 51 30.9 &   4.8 &   11 &  3.8 $\pm$ 1.1 &  0.8 & 0.73 \\
 35 & 12 22 49.70 & +15 47 43.7 &   4.4 &   10 &  2.3 $\pm$ 0.7 &  0.5 & 0.80 \\
 36 & 12 22 57.71 & +15 49 36.5 &   4.2 &    9 &  5.0 $\pm$ 1.7 &  1.0 & 0.67 \\
 37 & 12 22 45.94 & +15 47 16.6 &   4.1 &   11 &  2.8 $\pm$ 0.8 &  0.6 & 0.73 \\
 38 & 12 22 53.06 & +15 48 33.4 &   4.1 &   10 &  2.2 $\pm$ 0.7 &  0.5 & 0.60 \\
 39 & 12 23 05.67 & +15 48 53.1 &   4.1 &    9 &  2.0 $\pm$ 0.7 &  0.4 & 1.00 \\
 40 & 12 22 51.06 & +15 49 47.6 &   4.0 &   10 &  3.5 $\pm$ 1.1 &  0.7 & 0.70 \\
 41 & 12 22 57.65 & +15 48 39.4 &   3.9 &    9 &  2.0 $\pm$ 0.7 &  0.4 & 0.33 \\
 42 & 12 22 57.54 & +15 48 16.5 &   3.7 &    9 &  2.0 $\pm$ 0.7 &  0.4 & 0.89 \\
 43 & 12 22 54.89 & +15 50 16.7 &   3.6 &    8 &  2.6 $\pm$ 0.9 &  0.5 & 0.62 \\
 44 & 12 22 54.49 & +15 50 15.4 &   3.6 &    8 &  2.6 $\pm$ 0.9 &  0.5 & 0.12 \\
 45 & 12 23 03.02 & +15 48 27.0 &   3.6 &    8 &  1.8 $\pm$ 0.6 &  0.4 & 1.00 \\
 46 & 12 22 52.68 & +15 48 36.2 &   3.4 &    9 &  2.0 $\pm$ 0.7 &  0.4 & 0.44 \\
 47 & 12 22 48.08 & +15 50 53.5 &   3.3 &    8 &  2.7 $\pm$ 1.0 &  0.6 & 0.50 \\
 48 & 12 22 55.42 & +15 49 16.8 &   3.2 &   39 &  8.8 $\pm$ 1.4 &  1.8 & 0.31 & 2mass \\
 49 & 12 22 53.23 & +15 48 32.2 &   3.1 &    7 &  1.6 $\pm$ 0.6 &  0.3 & 0.43 \\
 50 & 12 23 05.62 & +15 50 44.5 &   3.0 &    6 &  2.0 $\pm$ 0.8 &  0.4 & 0.50 \\
\enddata
\tablecomments{The quantities are as defined for
Table~\ref{ngc4194src}.  Counterparts C1-C6 indicate sources from
\citet{Kaaret01}.}   \end{deluxetable}

\begin{deluxetable}{rllrrcccl}
\tabletypesize{\scriptsize}
\tablecaption{X-ray sources within the field of NGC 4945
  \label{ngc4945src}}
\tablewidth{0pt}
\tablehead{ \# & \colhead{RA} & \colhead{DEC} & 
  \colhead{S/N} & \colhead{Counts} &  
  \colhead{Flux} & \colhead{Luminosity} & \colhead{Hardness}  &
  \colhead{Counterpart} \\
	& & & & & ($10^{-15} \, \rm erg \, cm^{-2} \, s^{-1}$) & ($10^{38} \, \rm erg \, s^{-1}$) }
\startdata
  1 & 13 05 21.95 & -49 28 26.6 & 389.5 & 1349 & 466 $\pm$ 13  & 5.0   & 0.89  \\
  2 & 13 05 32.88 & -49 27 34.1 & 358.5 & 1311 & 439 $\pm$ 12  & 4.7   & 0.93  \\
  3 & 13 05 38.10 & -49 25 45.5 & 232.4 &  705 & 328 $\pm$ 12  & 3.5   & 0.91  \\
  4 & 13 05 18.54 & -49 28 24.0 & 189.0 &  496 & 172 $\pm$  8  & 1.85  & 0.82  \\
  5 & 13 05 35.49 & -49 29 11.4 & 139.7 &  340 & 118 $\pm$  6  & 1.27  & 0.82  \\
  6 & 13 05 34.61 & -49 27 51.8 & 124.9 &  297 &  99 $\pm$  6  & 1.06  & 0.92  \\
  7 & 13 05 27.49 & -49 28 05.2 &  86.6 &  341 & 121 $\pm$  7  & 1.29  & 1.00  \\
  8 & 13 05 22.25 & -49 29 12.5 &  86.5 &  223 &  76 $\pm$  5  & 0.8   & 0.66  \\
  9 & 13 05 22.86 & -49 29 01.4 &  74.5 &  197 &  67 $\pm$  5  & 0.7   & 0.96  \\
 10 & 13 05 25.47 & -49 28 32.4 &  70.4 &  194 &  66 $\pm$  5  & 0.7   & 0.98  \\
 11 & 13 05 40.78 & -49 26 03.6 &  62.5 &  115 &  54 $\pm$  5  & 0.6   & 0.99  \\
 12 & 13 05 21.18 & -49 27 41.3 &  58.0 &  157 &  59 $\pm$  5  & 0.6   & 0.82  \\
 13 & 13 05 28.96 & -49 29 44.3 &  57.7 &  133 &  46 $\pm$  4  & 0.5   & 0.86  \\
 14 & 13 05 16.36 & -49 27 34.8 &  54.3 &  128 &  44 $\pm$  4  & 0.5   & 0.80  \\
 15 & 13 05 21.69 & -49 27 37.0 &  49.6 &  130 &  46 $\pm$  4  & 0.5   & 0.77  \\
 16 & 13 04 42.57 & -49 34 49.5 &  44.4 &  350 & 196 $\pm$ 10  & 2.1   & 0.96  \\
 17 & 13 05 22.40 & -49 26 57.5 &  41.1 &   95 &  32 $\pm$  3  & 0.3   & 0.87  \\
 18 & 13 05 22.51 & -49 29 35.3 &  40.1 &   88 &  31 $\pm$  3  & 0.3   & 0.93  \\
 19 & 13 05 24.35 & -49 27 21.8 &  29.0 &   59 &  20 $\pm$  3  & 0.2   & 0.78  \\
 20 & 13 05 28.95 & -49 27 05.2 &  25.0 &   54 &  18 $\pm$  2  & 0.2   & 0.91  \\
 21 & 13 05 22.79 & -49 28 52.9 &  22.2 &   40 &  14 $\pm$  2  & 0.15  & 0.98  \\
 22 & 13 05 22.67 & -49 27 53.0 &  21.9 &   48 &  16 $\pm$  2  & 0.18  & 0.96  \\
 23 & 13 05 27.11 & -49 28 04.4 &  19.4 &   73 &  26 $\pm$  3  & 0.3   & 0.95  \\
 24 & 13 05 11.04 & -49 31 26.1 &  18.3 &   62 &  22 $\pm$  3  & 0.2   & 0.39  \\
 25 & 13 05 29.79 & -49 26 43.1 &  16.2 &   31 &  10 $\pm$  2  & 0.11  & 0.81  \\
 26 & 13 05 09.79 & -49 31 42.8 &  16.0 &   62 &  21 $\pm$  3  & 0.2   & 0.95  \\
 27 & 13 05 37.23 & -49 26 02.0 &  15.0 &   29 &  13 $\pm$  2  & 0.14  & 0.86  \\
 28 & 13 05 25.42 & -49 28 24.0 &  13.1 &   33 &  11 $\pm$  2  & 0.12  & 1.00  \\
 29 & 13 05 13.26 & -49 28 33.7 &  10.9 &   24 & 8.7 $\pm$ 1.7 & 0.09  & 0.75  \\
 30 & 13 05 43.39 & -49 24 31.4 &   8.8 &   20 &  10 $\pm$  2  & 0.11  & 0.95  \\
 31 & 13 05 35.49 & -49 25 27.0 &   8.4 &   16 & 7.5 $\pm$ 1.9 & 0.08  & 1.00  \\
 32 & 13 05 26.65 & -49 27 41.9 &   7.0 &   21 & 7.1 $\pm$ 1.5 & 0.08  & 0.62  \\
 33 & 13 04 56.68 & -49 33 40.2 &   6.2 &   33 &  17 $\pm$  3  & 0.18  & 0.82  \\
 34 & 13 05 30.01 & -49 28 20.2 &   5.9 &   13 & 4.5 $\pm$ 1.3 & 0.05  & 0.85  \\
 35 & 13 05 24.91 & -49 28 39.9 &   5.3 &   11 & 3.8 $\pm$ 1.1 & 0.04  & 1.00  \\
 36 & 13 05 39.10 & -49 25 29.7 &   5.1 &   11 & 5.1 $\pm$ 1.6 & 0.06  & 0.82  \\
 37 & 13 05 11.13 & -49 29 02.5 &   4.9 &   10 & 3.5 $\pm$ 1.1 & 0.04  & 1.00  \\
 38 & 13 05 33.16 & -49 27 01.4 &   4.9 &    9 & 3.0 $\pm$ 1.0 & 0.03  & 0.44  \\
 39 & 13 05 27.47 & -49 29 30.8 &   4.9 &   10 & 3.5 $\pm$ 1.1 & 0.04  & 1.00  \\
 40 & 13 05 31.19 & -49 27 49.0 &   4.6 &   10 & 3.4 $\pm$ 1.1 & 0.04  & 0.90  \\
 41 & 13 05 31.78 & -49 26 58.2 &   4.4 &   10 & 3.4 $\pm$ 1.1 & 0.04  & 0.80  \\
 42 & 13 05 35.63 & -49 26 58.8 &   4.2 &    7 & 2.4 $\pm$ 0.9 & 0.03  & 1.00  \\
 43 & 13 05 23.14 & -49 27 54.9 &   4.1 &   11 & 3.7 $\pm$ 1.1 & 0.04  & 0.73  \\
 44 & 13 05 36.70 & -49 27 22.1 &   4.0 &    9 & 3.0 $\pm$ 1.0 & 0.03  & 1.00  \\
 45 & 13 05 14.92 & -49 30 42.5 &   3.8 &   12 & 4.0 $\pm$ 1.2 & 0.04  & 0.50  \\
 46 & 13 05 52.86 & -49 23 12.2 &   3.7 &   11 & 5.5 $\pm$ 1.7 & 0.06  & 0.45  \\
 47 & 13 05 42.95 & -49 25 09.1 &   3.6 &    7 & 3.3 $\pm$ 1.3 & 0.04  & 1.00  \\
 48 & 13 05 19.53 & -49 30 03.2 &   3.0 &    5 & 1.7 $\pm$ 0.8 & 0.02  & 0.60  \\
\enddata
\tablecomments{The quantities are as defined for
Table~\ref{ngc4194src}.}   \end{deluxetable}

\label{lastpage}

\end{document}